\documentclass[twocolumn,preprintnumbers]{revtex4}

\usepackage{graphicx}
\usepackage{epsf}
\usepackage{amsmath,amssymb}
\newcommand{\bvec}{\boldsymbol}

\begin{document}
\title{Cluster and toroidal aspects of isoscalar dipole excitations in $^{12}$C}
\author{Yoshiko Kanada-En'yo}
\affiliation{Department of Physics, Kyoto University, Kyoto 606-8502, Japan}
\author{Yuki Shikata}
\affiliation{Department of Physics, Kyoto University, Kyoto 606-8502, Japan}
\author{Horiyuki Morita}
\affiliation{Department of Physics, Kyoto University, Kyoto 606-8502, Japan}
\begin{abstract}
We investigated cluster and toroidal aspects of isoscalar dipole
excitations in $^{12}$C based on the shifted basis antisymmetrized molecular dynamics 
combined with the generator coordinate method, which can describes 1p-1h excitations and
$3\alpha$ dynamics.
In the $E=10-15$ MeV region, 
we found two low-energy dipole modes separating from the giant dipole resonance. 
One is the developed $3\alpha$-cluster state and the other is the toroidal dipole mode. 
The cluster state is characterized by the large amplitude cluster motion beyond the 
1p-1h model space, whereas
the toroidal dipole mode is predominantly described by 1p-1h excitations on the 
ground state. The low-energy dipole states are remarkably excited by  
the toroidal dipole operator, which can measure the nuclear vorticity. 
For compressive dipole transition strengths, a major part is distributed in the $30-50$ MeV region 
for the giant dipole resonance, and
5\% of the total energy weighted 
sum exist in the $E<20$ MeV region.
\end{abstract}
\maketitle

\section{Introduction}
In these decades, researches of isoscalar monopole (ISM) and dipole (ISD) excitations 
have been  proceeding remarkably
through experiments with $\alpha$ inelastic scattering. 
Recently, particular attention has been paid to low-energy (LE) monopole and dipole strengths 
below the giant resonances (GR). 
Since the ISM and ISD operators corresponding to 
compressive modes can directly excite inter-cluster motions, 
they are good probes for cluster states as discussed by Yamada {\it et al.} \cite{Yamada:2011ri}
and Chiba {\it et al.} \cite{Chiba:2015khu}. 
Indeed,  in such nuclei as $^{16}$O and $^{24}$Mg, 
the LE- ISM and ISD strengths are described by 
cluster states, which appear separating from the collective modes of the GRs. 

In the progress in physics of unstable nuclei, the LE-ISD excitations have been also discussed 
in relation to isovector (IV) dipole excitations (see, e.g.,  reviews in 
Refs.~\cite{Paar:2007bk,aumann-rev,Savran:2013bha,Bracco:2015hca} and references therein). 
Owing to experimental studies with hadronic probes, 
information of isospin characters of the LE dipole (LED) excitations are becoming
available for various nuclei.
The IS giant dipole resonances (GDR), which correspond to 
the collective compressive dipole mode, 
have been observed for stable nuclei in the energy region higher than that of 
the IVGDR for the proton-neutron opposite oscillation mode.
Below the ISGDR, significant LE-ISD 
strengths have been known in stable nuclei 
\cite{Harakeh:1981zz,Decowski:1981pcz,Poelhekken:1992gvp}. 
In the ISD strengths of $^{16}$O and $^{40}$Ca, $4-5\%$ 
of the energy weighted sum rule have been observed in the $E\le 10$ MeV region. 

In order to understand the LED strengths,
the toroidal dipole (TD) mode (called also the torus or vortical mode)  has been proposed
with hydrodynamical models \cite{semenko81,Ravenhall:1987thb}.
The TD mode carries vorticity and its character is much 
different from the compressive dipole (CD) mode, which corresponds to the
ordinary ISD mode contributing to the ISGDR. The energy of the TD mode is expected to be lower 
than the ISGDR energy because it conserves the nuclear density.
In these years, microscopic calculations with quasiparticle phonon model (QPM) and random phase approximation (RPA) 
have been achieved, and toroidal natures of the LED excitations
in various nuclei have been investigated
\cite{Paar:2007bk,Vretenar:2001te,Ryezayeva:2002zz,Papakonstantinou:2010ja,Kvasil:2011yk,Repko:2012rj,Nesterenko:2016qiw}. 

In the ISD stregntsh of $^{12}$C measured by 
$\alpha$ inelastic scattering, 
several percentages of the energy weighted sum rule
have been observed 
in the $E\le 20$ MeV region below the ISGDR energy \cite{John:2003ke}. 
In the LE-ISD strengths, there is a peak for the $1^-_1$ state at 10.8 MeV. In addition, 
another peak (or bump) structure
around 15 MeV exists indicating a possible existence of the LED mode other than the $1^-$ (10.8 MeV). 
In theoretical studies of $^{12}$C, a variety of $3\alpha$-cluster states have been suggested in excited states 
above the $3\alpha$ threshold energy (7.16 MeV). 
Microscopic $3\alpha$-cluster models \cite{kamimura-RGM1,kamimura-RGM2,uegaki1,uegaki3} describe 
a spatially developed $3\alpha$-cluster structure of the $1^-_1$ state. 
Even though cluster models are useful for $3\alpha$-cluster states of $^{12}$C, the models 
{\it a priori} assume three $\alpha$ clusters, and therefore 
they are insufficient to study 1p-1h excitations and unable to describe 
GRs.
In order to take into account coherent 1p-1h excitations for the GRs 
as well as the large amplitude cluster modes, one of the authors, Y. K-E.,  has extended the antisymmetrized molecular dynamics (AMD) 
\cite{KanadaEnyo:1995tb,KanadaEnyo:1995ir,KanadaEn'yo:2001qw,KanadaEn'yo:2012bj} 
and constructed a new method, 
the shifted basis AMD combined 
with the generator coordinate method (GCM) with respect to the inter-cluster motion, which we 
named ``sAMD+GCM''  \cite{Kanada-En'yo:2013dma,Kanada-Enyo:2015knx,Kanada-Enyo:2015vwc}. 
In the previous work \cite{Kanada-Enyo:2015vwc}, we applied 
the sAMD+GCM to calculate ISM and ISD transition strengths of $^{12}$C 
and obtained significant LE-ISD strengths in the $E=10-15$ MeV region
well separating from the high-energy strengths for the ISGDR. 

Our aim in this paper is to clarify natures of the LE-ISD excitations of $^{12}$C, 
in particular, cluster and toroidal aspects, by reanalysis of the previous calculation. 
To probe the toroidal nature, we adopt the TD operator
in addition to the CD operator for the ordinary ISD and show the remarkable 
TD strengths for two LE-ISD modes, the $1^-_1$ with the developed $3\alpha$-cluster structure and 
the $1^-_2$ dominated by 1p-1h configurations with 
significant cluster breaking. We also discuss the TD mode
from a cluster picture and its connection with 1p-1h excitations 
in a shell-model limit. 

The paper is organized as follows. 
The framework of the sAMD+GCM for $^{12}$C is explained in Sec.~\ref{sec:framework}.
Definitions of ISD operators are given in Sec.~\ref{sec:dipole}.
Section  \ref{sec:results} shows 
the calculated results and discusses properties of the LED modes.
The intrinsic structure of the TD mode and its shell-model limit are presented 
in Sec.~\ref{sec:discussions}.
Finally, the paper concludes with a summary and an outlook  in section \ref{sec:summary}.

\section{sAMD+GCM}\label{sec:framework}
The sAMD+GCM has been applied to 
ISM, ISD, and $E1$ excitations of light nuclei such as $^{10}$Be, $^{12}$C and  $^{16}$O
\cite{Kanada-En'yo:2013dma,Kanada-Enyo:2015knx,Kanada-Enyo:2015vwc}.
A similar method has been recently applied to 
$E1$ and ISD excitations of $^{26}$Ne by Kimura \cite{Kimura:2016heo}.
In the previous work, 
we applied the sAMD+GCM to $^{12}$C by taking into account 
$3\alpha$-cluster configurations.
For the detailed procedure of the sAMD+GCM calculation,
the reader is referred to Ref.~\cite{Kanada-Enyo:2015vwc}.

In the AMD method, a basis wave function is given by a Slater determinant,
\begin{equation}
 \Phi_{\rm AMD}({\bvec{Z}}) = \frac{1}{\sqrt{A!}} {\cal{A}} \{
  \varphi_1,\varphi_2,...,\varphi_A \},\label{eq:slater}
\end{equation}
where  ${\cal{A}}$ is the antisymmetrizer, and  $\varphi_i$ is 
the $i$th single-particle wave function written by a product of
spatial, spin, and isospin
wave functions, 
\begin{eqnarray}
 \varphi_i&=& \phi_{{\bvec{X}}_i}\chi_i\tau_i,\\
 \phi_{{\bvec{X}}_i}({\bvec{r}}_j) & = &  \left(\frac{2\nu}{\pi}\right)^{3/4}
\exp\bigl[-\nu({\bvec{r}}_j-\bvec{X}_i)^2\bigr],
\label{eq:spatial}\\
 \chi_i &=& (\frac{1}{2}+\xi_i)\chi_{\uparrow}
 + (\frac{1}{2}-\xi_i)\chi_{\downarrow},
\end{eqnarray}
where $\phi_{{\bvec{X}}_i}$ and $\chi_i$ are the spatial and spin functions, respectively, and 
$\tau_i$ is the isospin
function fixed to be proton or neutron. The width parameter $\nu=0.19$ fm$^{-2}$ is used 
to minimize the ground state energy of $^{12}$C. 
The AMD wave function
is specified by a set of variational parameters, ${\bvec{Z}}\equiv 
\{{\bvec{X}}_1,\ldots, {\bvec{X}}_A,\xi_1,\ldots,\xi_A \}$, which indicate
Gaussian centroids and spin orientations 
of all single-particle wave functions.

In order to obtain 
the ground state wave function,  
the variation after 
angular-momentum and parity projections (VAP) for the $0^+$-projected AMD wave function is performed as  
\begin{eqnarray}
&& \frac{\delta}{\delta{\bvec{X}}_i}
\frac{\langle \Phi|H|\Phi\rangle}{\langle \Phi|\Phi\rangle}=0,\\
&& \frac{\delta}{\delta\xi_i}
\frac{\langle \Phi|H|\Phi\rangle}{\langle \Phi|\Phi\rangle}=0,\\
&&\Phi= P^{J\pi}_{MK}\Phi_{\rm AMD}({\bvec{Z}}),
\end{eqnarray}
where $P^{J\pi}_{MK}$ is the angular-momentum and parity projection operator. 
We denote the optimized parameter set $\bvec{Z}$ 
for the ground state as $\bvec{Z}^0_\textrm{VAP}=\{\bvec{X}^0_1,\ldots,\xi^0_{1},\ldots\}$.

To take into account 1p-1h excitations on the top of the obtained ground state, 
we consider small variation of single-particle wave functions of  
$\Phi_{\rm AMD}({\bvec{Z}^0_\textrm{VAP}})$ by shifting the Gaussian centroid 
of the $i$th single-particle wave function,
${\bvec{X}}^0_i\rightarrow {\bvec{X}}^0_i+\epsilon{\bvec{e}}_\sigma$, where
$\epsilon$ is an enough small constant and 
${\bvec{e}}_\sigma$ ($\sigma=1,\ldots,8$) are unit vectors for 8 directions 
explained in Ref.~\cite{Kanada-Enyo:2015vwc}.
As for the intrinsic spin of the shifted single-particle wave function, 
the spin non-flip and flip states 
are adopted. Consequently, 
totally $16A=192$ wave functions 
of the spin non-flip and flip shifted AMD wave functions are superposed. We call this method, 
the shifted basis AMD (sAMD).

\begin{figure}[!h]
\begin{center}
\includegraphics[width=4cm]{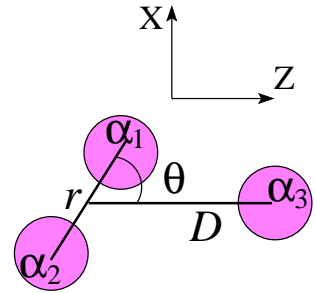} 	
\end{center}
  \caption{(color online) Schematic figure of $3\alpha$ configurations in 
the GCM for $^{12}\textrm{C}$.
\label{fig:3-alpha}}
\end{figure}

In addition to
1p-1h excitations expressed by the sAMD, 
we combine the GCM with the sAMD 
in order to take into account large amplitude dynamics of three $\alpha$ clusters by 
superposing various $3\alpha$ configurations 
written by the Brink-Bloch cluster model wave functions \cite{Brink66}. 
Practically, a $3\alpha$-cluster configuration can be expressed with the
AMD wave function
by setting the parameters as $\bvec{X}_i=\bvec{S}_k$
for four nucleons ($p\uparrow$, $p\downarrow$, $n\uparrow$, and $n\downarrow$) in the $k$th
$\alpha$ cluster ($\alpha_k$). 
$\bvec{S}_k$ indicates the $\alpha_k$ cluster position and is given as 
\begin{eqnarray}
\bvec{S}_1&=& -\frac{1}{3}\bvec{D}+ \frac{\bvec{r}}{2}, \\
\bvec{S}_2&=& -\frac{1}{3}\bvec{D}- \frac{\bvec{r}}{2} ,\\
\bvec{S}_3&=& \frac{2}{3}\bvec{D}.
\end{eqnarray}
Here the $\alpha_1$-$\alpha_2$ and 
$\alpha_3$-$\alpha_1\alpha_2$ relative vectors, 
$\bvec{r}$ and $\bvec{D}$,  are chosen to be 
$\bvec{r}=(r\cos\theta,r\sin\theta,0)$ and $\bvec{D}=(0,0,D)$, respectively. 
$\theta$ is  
the angle between two vectors as shown in the schematic in Fig.~\ref{fig:3-alpha}.
We use $r=\{0.8$, 1.8, \ldots, 4.8 fm\}, 
$D=$\{1, 2, \ldots, 7 fm\}, and $\theta=(\pi/8) i$ $ (i=0,\ldots,4)$. Finally, 
in the sAMD+GCM, we  
superpose all the basis wave functions of the sAMD and $3\alpha$ wave functions 
in addition to the original VAP wave function ($\Phi_{\rm AMD}({\bvec{Z}^0_\textrm{VAP}})$)  
as done in the previous work \cite{Kanada-Enyo:2015vwc}. 
The final wave functions  $\Psi(J^\pi_k)$ for the $0^+_k$ and $1^-_k$ states
are expressed by superposition of the $J^\pi$-projected wave functions 
with coefficients determined by diagonalizing the norm and Hamiltonian 
matrices.

In the sAMD+GCM method, 
the ground state is obtained by the VAP, and therefore, it already contains 
the ground state correlation beyond mean-field approximation. In the sAMD, 
1p-1h excitations on the ground state are taken into account 
by the linear combination of shifted Gaussian wave packets, and 
large amplitude cluster motion is treated by superposition of $3\alpha$-cluster wave functions 
in the GCM. In the present method, the center of mass motion is exactly removed. 

\section{ISD operators and transition strengths}\label{sec:dipole}

In order to probe vorticity of the nuclear current, 
two kinds of operators have been used. One is the ``toroidal mode'' 
originally determined by the second order correction 
in the long-wave approximation of the transition $E\lambda$ operator in an
electromagnetic field \cite{semenko81,Dubovik75}, 
and the other is the ``vortical mode'' in Ravenhall-Wambach's prescription \cite{Ravenhall:1987thb}.
Kvasil  {\it et al.} 
described general treatment of toroidal, compressive, and vortical modes and their
relation to each other \cite{Kvasil:2011yk}. They showed that the toroidal mode 
is a good probe for LE-ISD excitations rather than the vortical mode, though both the 
TD and vortical dipole (VD) operators can measure the vorticity. 
In this paper, we mainly discuss the TD and CD transitions. 
We also show the VD transitions just for comparison.

We use the TD, CD, and VD operators 
defined in Ref.~\cite{Kvasil:2011yk} as 
\begin{eqnarray}
M_\textrm{TD}(\mu)&=&\frac{-i}{2\sqrt{3}c}\int d\bvec{r} \bvec{j}(\bvec{r}) \nonumber\\
&\times& 
\left [
\frac{\sqrt{2}}{5} r^2 
\bvec{Y}_{12\mu}(\hat{\bvec{r}})+
r^2 \bvec{Y}_{10\mu} (\hat{\bvec{r}})  
\right ],\\
M_\textrm{CD}(\mu)&=&\frac{-i}{2\sqrt{3}c}\int d\bvec{r} \bvec{j}(\bvec{r}) 
 \nonumber\\
&\times& 
\left [  \frac{2\sqrt{2}}{5} r^2 \bvec{Y}_{12\mu}(\hat{\bvec{r}}) - r^2 \bvec{Y}_{10\mu} (\hat{\bvec{r}}) 
\right ],\\
M_\textrm{VD}(\mu)&=&\frac{-i}{2\sqrt{3}c}\int d\bvec{r} \bvec{j}(\bvec{r}) 
 \nonumber\\
&\times&
\left [ \frac{3\sqrt{2}}{5} r^2 
\bvec{Y}_{12\mu}(\hat{\bvec{r}})\right ],
\end{eqnarray}
where $\bvec{j}(\bvec{r})$ is the current density operator and 
$\bvec{Y}_{\lambda L\mu}$ is the vector spherical  harmonics.
Using the dipole operators we analyze the toroidal nature of the dipole excitations 
in the same matter as done in Ref.~\cite{Kanada-Enyo:2017uzz}, which discussed 
dipole excitations of $^{10}$Be. 
The detailed definition of $\bvec{j}(\bvec{r})$ as well as that of the density operator $\rho(\bvec{r})$ is
explained in appendix of Ref.~\cite{Kanada-Enyo:2017uzz}.
For the current density, we take into account only the convection part of the 
nuclear current but omit its magnetization (spin) part.
The matrix elements for the dipole transitions, $|0^+_1\rangle \to |1^-_k\rangle$,
are given with the transition current density
$\delta\bvec{j}(\bvec{r})\equiv \langle 1^-_k\ | \bvec{j}(\bvec{r})|0^+_1 \rangle$  
as 
\begin{eqnarray}
&&\langle  1^-_k|M_\textrm{TD}(\mu) |0^+_1  \rangle =\nonumber\\
&&\frac{-i}{2\sqrt{3}c}\int d\bvec{r} \delta\bvec{j}(\bvec{r})
\left [
\frac{\sqrt{2}}{5} r^2 
\bvec{Y}_{12\mu}(\hat{\bvec{r}})+
r^2 \bvec{Y}_{10\mu} (\hat{\bvec{r}})  
\right ],\\
&&\langle 1^-_k|M_\textrm{CD}(\mu) |0^+_1  \rangle=\nonumber\\
&&\frac{-i}{2\sqrt{3}c}\int d\bvec{r} \delta\bvec{j}(\bvec{r}) 
\left [  \frac{2\sqrt{2}}{5} r^2 \bvec{Y}_{12\mu}(\hat{\bvec{r}}) - r^2 \bvec{Y}_{10\mu} (\hat{\bvec{r}}) 
\right ],\\
&&\langle  1^-_k |M_\textrm{VD}(\mu) |0^+_1  \rangle=\nonumber\\
&&\frac{-i}{2\sqrt{3}c}\int d\bvec{r} \delta\bvec{j}(\bvec{r})
\left[\frac{3\sqrt{2}}{5} r^2 
\bvec{Y}_{12\mu}(\hat{\bvec{r}})\right].
\end{eqnarray}
Note that the CD matrix element can be transformed to 
the ordinary ISD matrix element (labeled IS1) by using the continuity equation 
$\nabla\cdot  \bvec{j} = - \frac{i}{\hbar} \left [H,\rho \right]$ as 
\begin{eqnarray}
\langle 1^-_k|M_\textrm{CD}(\mu)|  0^+_1 \rangle &=&
\frac{E}{10\hbar c} \langle 1^-_k |M_{IS1}(\mu)|  0^+_1\rangle, \\
M_\textrm{IS1}(\mu)&\equiv&\int d\bvec{r} \rho(\bvec{r}) r^3 Y_{1\mu} (\hat{\bvec{r}}),
\end{eqnarray}
where $E$ is the excitation energy of the $1^-_k$.  

For the $0^+_1$ and $1^-_k$ wave functions obtained by the sAMD+GCM, 
we calculate the transition strengths of the dipole operators 
\begin{eqnarray}
&&\tilde B(\textrm{TD,CD,VD}; 0^+_1 \to 1^-_k) \nonumber \\
&&\equiv  \left(\frac{10\hbar c}{E}\right )^2 \left| \langle 1^-_k||M_\textrm{TD,CD,VD} ||0^+_1 \rangle \right |^2.
\end{eqnarray}
Here we define
the scaled transition strengths with the factor $\left(\frac{10\hbar c}{E}\right )^2$
so that $\tilde B(\textrm{CD})$ corresponds to the IS1 strength 
$B(\textrm{IS1})= \left| \langle 1^-_k||M_\textrm{IS1} ||0^+_1 \rangle \right |^2$.

\section{Results}\label{sec:results}

\subsection{Effective interactions}
The adopted effective interactions are the same as those used in the previous work 
\cite{Kanada-Enyo:2015vwc}.
The central force is the MV1 force \cite{TOHSAKI} consisting of 
two-range Gaussian two-body terms and a zero-range three-body term.
As for the parametrization of the MV1 force,  
the case 1 with the Bartlett, Heisenberg, and Majorana parameters, $b=h=0$ and $m=0.62$, is used. 
In addition to the central force, the two-range Gaussian spin-orbit term of the 
G3RS force \cite{LS1,LS2}  with he strengths 
$u_{I}=-u_{II}=3000$ MeV is supplemented.
This set of interaction parameters describes well properties of the ground and excited states 
of $^{10}$Be and $^{12}$C in the AMD+VAP calculations
\cite{KanadaEn'yo:1998rf,KanadaEn'yo:1999ub,KanadaEn'yo:2006ze}.

\section{ISD excitations of $^{12}$C obtained with sAMD+GCM}

\begin{figure}[!h]
\begin{center}
\includegraphics[width=7.5cm]{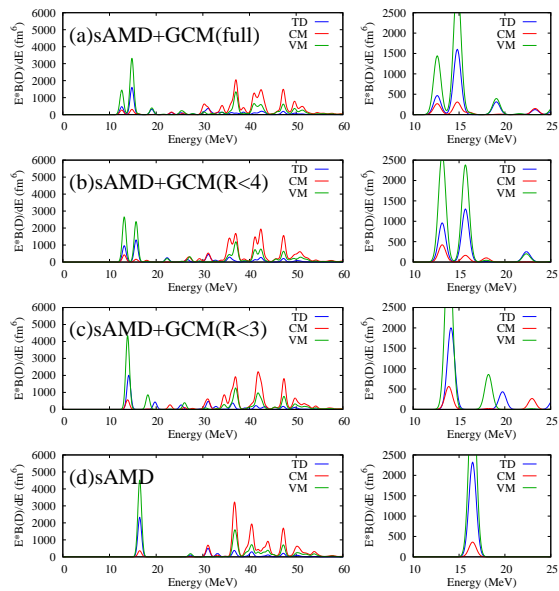} 	
\end{center}
  \caption{(color online) The strength functions of the TD, CD, and VD transitions calculated with 
(a) the full sAMD+GCM, 
(b) sAMD with $|\bvec{R}_k| \le 4$ fm $3\alpha$ configurations, 
(c) that with $|\bvec{R}_k| \le 3$ fm $3\alpha$ configurations, 
and (d) sAMD without $3\alpha$ configurations. 
The strengths of discrete states are smeared by Gaussian with the range 
$\gamma=1/\sqrt{\pi}$ MeV. 
\label{fig:c12-isd}}
\end{figure}

We calculate the TD, CD, and VD transitions  from the ground 
to $1^-$ states obtained by the sAMD+GCM. The calculated 
transition strengths are shown in Fig.~\ref{fig:c12-isd} (a).
The remarkable TD and VD strengths are obtained for LED states in the $E=10-15$ MeV region,
whereas the CD operator strengths are mainly distributed in the 
high-energy part corresponding to the ISGDR. 
The LE strengths are concentrated on two dipole states; 
one is the $1^-_1$ state at 12.6 MeV, which we assign to  the experimental $1^-$(10.844  MeV) state, 
and the other is the $1^-_2$ state at 14.8 MeV. 
The $1^-_2$ state has a significant 1p-1h component and approximately 
described within the sAMD model space.
On the other hand, the $1^-_1$ state has a spatially developed $3\alpha$-cluster structure, which 
is a large amplitude cluster mode beyond the sAMD model space. 
It means that,  in the $1^-$ spectra of $^{12}$C, the large amplitude cluster mode 
exists in the energy lower than the 1p-1h dominant dipole excitation.
The $3\alpha$-cluster structure of the $1^-_1$ state
is consistent with the $3\alpha$-GCM calculation of Refs.~\cite{uegaki1,uegaki3}.

In order to see the significance of the large amplitude cluster motion in the LE-ISD excitations, 
we perform calculations within truncated model spaces
by reducing $3\alpha$ configurations.
Figures \ref{fig:c12-isd}(b), (c), and (d) show the ISD strengths obtained by 
the truncated calculations using the sAMD with $|\bvec{S}_k| \le 4$ fm
$3\alpha$ configurations, that with $|\bvec{S}_k| \le 3$ fm,  and 
the sAMD without $3\alpha$ configurations, respectively. Note that 
$|\bvec{S}_k|$ is the distance of  the $\alpha_k$-cluster center from the origin. 
In the calculations, the initial state is fixed to be the $0^+_1$ state obtained by the 
full sAMD+GCM. 
As seen in Fig.~\ref{fig:c12-isd}(d), the sAMD without $3\alpha$ configurations shows 
only one $1^-$ state in $E<20$ MeV. 
The $1^-$ state has the dominant 1p-1h component with the remarkable TD strength, 
and approximately corresponds to the 
$1^-_2$ state of the full sAMD+GCM. In the sAMD result, there is no 
low-lying  $3\alpha$-cluster state corresponding to the $1^-_1$ of the full sAMD+GCM. 
As $3\alpha$-cluster configurations are added to the sAMD model space, the $3\alpha$-cluster state
appears around 20 MeV in the sAMD+GCM($|\bvec{S}_k| \le 3$ fm)  (Fig.~\ref{fig:c12-isd}(c)), comes down 
to 15 MeV  in the sAMD+GCM($|\bvec{S}_k| \le 4$ fm)  (Fig.~\ref{fig:c12-isd}(b)), and finally becomes 
lower than the 1p-1h dominant $1^-_2$ state 
in the full sAMD+GCM. In other words, 
as the large amplitude inter-cluster motion develops,  
the cluster mode 
comes down to the lowest $1^-$ state crossing the 1p-1h state.

\begin{figure}[!h]
\begin{center}
\includegraphics[width=6.5cm]{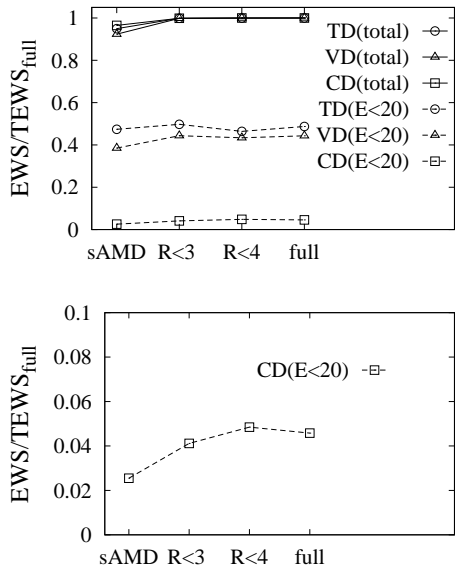} 	
\end{center}
  \caption{Energy-weighted TD, VD, CD strengths summed  
up to $E=20$ MeV obtained by the sAMD,
sAMD+GCM ($|\bvec{R}_k| \le 3$ fm),
sAMD+GCM ($|\bvec{R}_k| \le 4$ fm), and 
full sAMD+GCM calculations.
The ratio to the total energy weighted sum (TEWS$_\textrm{full}$) value of  
the full sAMD+GCM is shown.
The TEWS value of each calculation relative to the TEWS$_\textrm{full}$
is also shown. 
\label{fig:ews-ratio}}
\end{figure}

Figure \ref{fig:ews-ratio} shows the energy-weighted sum (EWS) of the 
TD, VD, CD strengths up to $E=20$ MeV 
obtained by the truncated and full calculations. 
The ratio to the  total energy-weighted sum (TEWS$_\textrm{full}$)
of the full sAMD+GCM is shown.
The EWS value of the LE TD (VD) strengths is as much as $40-50$\% of the TEWS 
almost independently from the truncation of the $3\alpha$ configurations. 
It means that the sum of the LE TD strengths is approximately described 
by the 1p-1h configurations within the sAMD  model space.
As already shown in Figs.~\ref{fig:c12-isd}(a) and (d), 
the strength of the LE TD transition is originally 
concentrated at the single peak in the sAMD, and it is split into two LED states 
in the full sAMD+GCM as a result of coupling with the large amplitude $3\alpha$-cluster mode.
By contrast, the LE CD transition strengths have only tiny percentages of the TEWS.
The ratio is 2.5\% in the sAMD, and it is raised up to $4-5$\% by $3\alpha$ configurations. 
It means that about a half of the LE CD strengths is contributed by
1p-1h configurations and another half comes from $3\alpha$ configurations.

\begin{figure}[htb]
\begin{center}
\includegraphics[width=7.5cm]{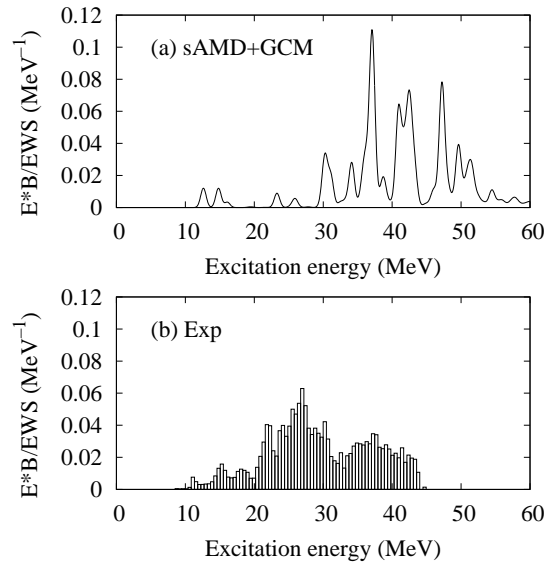} 	
\end{center}
  \caption{The EWSR ratio of the energy weighted IS1 strengths.
(a) The calculated strengths smeared by a Gaussian with the width $\gamma=1/\sqrt{\pi}$ MeV. 
(b) The experimental data measured by $\alpha$ inelastic scattering 
from Ref.~\cite{John:2003ke}. The figures are from Ref.~\cite{Kanada-Enyo:2015vwc}.
\label{fig:c12-ewsr}}
\end{figure}

Figure~\ref{fig:c12-ewsr} shows comparison of 
the calculated IS1 strengths  with the
experimental data measured by $\alpha$ inelastic scattering.  
The calculated LE IS1 strengths are comparable to the experimental data. 
We assign the lowest peak for the calculated $1^-_1$ state to the experimental 
IS1 strength for the $1^-_1$ (10.84 MeV).  
The bump structure around $E=15$ MeV in the experimental strength function
is a candidate for the calculated $1^-_2$ state.
In the present calculation, high-energy strengths for 
the ISGDR are distributed in the $E=30-50$ MeV region more or less 
higher than the experimental ISGDR energy. 

\section{Cluster and toroidal natures of the low-energy dipole excitations}\label{sec:discussions}

In the present calculation,  two LED modes are obtained in $E=10-15$ MeV. 
One is the large-amplitude $3\alpha$-cluster mode, and the other 
is the 1p-1h dominant dipole excitation.
We here discuss the cluster and toroidal natures of the LED states. 

\subsection{Occupation probabilities in shell-model expansion}

\begin{figure}[!h]
\begin{center}
\includegraphics[width=7.5cm]{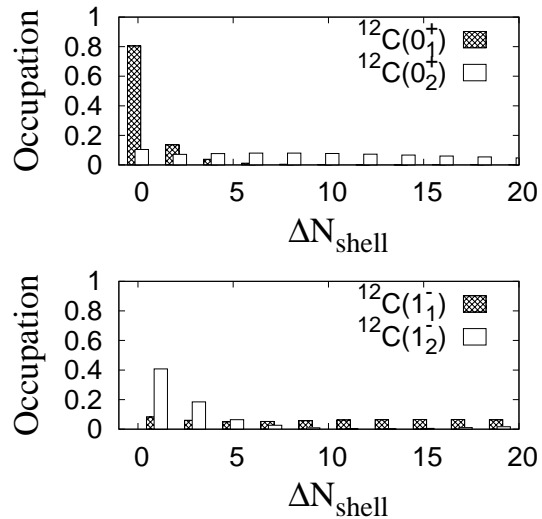} 	
\end{center}
  \caption{Occupation probability of harmonic oscillator quanta $N$ 
in the shell-model basis expansion of the $0^+_{1,2}$ and $1^-_{1,2}$ states
obtained by the sAMD+GCM calculation. The horizontal axis indicates the difference $\Delta N$ from the minimum quanta $N_\textrm{min}=8$.  
\label{fig:ho}}
\end{figure}

Figure \ref{fig:ho} shows the occupation probability of harmonic oscillator quanta $N$ 
in the shell-model basis expansion with the size parameter 
$b=1/\sqrt{2\nu}$.
The dominant component of the $0^+_1$ is the $0\hbar\omega$ configuration 
with 20\% mixing of higher shell configurations. By contrast, the occupation probability of the  
$0^+_2$ 
is distributed broadly in the high shell region because of the spatial developed 
$3\alpha$-cluster structure. Similarly to the $0^+_2$ state, the $1^-_1$ state shows very broad 
distribution of the occupation probability due to the developed cluster structure. 
The $1^-_2$ state contains 
40\% $1\hbar\omega$ component indicating the dominant 1p-1h excitations 
with significant higher-shell mixing. 

\subsection{Intrinsic structures}
In the sAMD+GCM calculation, each state is expressed by 
superposition of many different configurations of the shifted AMD and 3$\alpha$ wave functions. 
For intuitive understanding of the LED excitations, 
it is useful to consider a single Slater determinant which approximately describes 
the state of interest. 
We here perform simple model analysis by introducing model 
wave functions that have large overlap with the obtained $0^+_1$, $1^-_1$, and
$1^-_2$ states in order to discuss cluster and toroidal natures of the LED modes
in the intrinsic frame.

As for the model wave functions in the present analysis, we adopt an
extended $3\alpha$  (E-$3\alpha$) model  based on the quasi $\alpha$-cluster model 
proposed by Itagaki {\it et al.} \cite{Itagaki:2005sy}, in which the $\alpha$-breaking 
is taken into account by the cluster breaking parameter $\Lambda$. 
We start from the Brink-Bloch $3\alpha$-cluster wave functions, and incorporate 
the breaking of the twp $\alpha$ clusters, $\alpha_1$ and $\alpha_2$.

Let us first consider the $\theta=\pi/2$ case for
an isosceles triangle $3\alpha$ configuration. 
An E-$3\alpha$ wave function can be expressed by the AMD wave function with 
parameters
\begin{eqnarray}
\bvec{X}_i&=&\bvec{S}_1+ i\bvec{W}_i \quad (i=1,\cdots,4), \\
\bvec{X}_i&=&\bvec{S}_2+ i\bvec{W}_i  \quad (i=5,\cdots,8), \\
\bvec{X}_i&=&\bvec{S}_3 \quad (i=9,\cdots,12), \\
\bvec{W}_i&=&\Lambda \left( \bvec{e}_{\sigma i}\times \frac{\bvec{r}}{2}\right) \quad (i=1,\cdots,4), \\
\bvec{W}_i&=&\Lambda \left( \bvec{e}_{\sigma i}\times (-\frac{\bvec{r}}{2})\right) \quad (i=5,\cdots,8).
\end{eqnarray}
The spin and isospin functions are fixed to be 
$\tau_i\chi_i=p\uparrow, p\downarrow, n\uparrow, n\downarrow$ for $i=\{1,5,9\}$, $i=\{2,6,10\}$, $i=\{3,7,11\}$, $i=\{4,8,12\}$, respectively, and $\bvec{e}_{\sigma i}$ indicates the unit vector for the spin 
orientation.
$\Lambda$ describes the $\alpha$-breaking because of the spin-orbit interaction
through the imaginary part $\bvec{W}_i$ of the Gaussian centroid $\bvec{X}_i$, 
which depends on the intrinsic spin orientation. 
In the $\Lambda >0 $ case, $\bvec{W}_i$ indicates 
finite momenta of nucleons boosted by the spin-orbit potential. 
In the $\Lambda=0$ case,  
the E-$3\alpha$ wave function becomes equivalent to the ideal 
Brink-Bloch $3\alpha$-cluster wave function without the $\alpha$-breaking.
Next, a non-isosceles triangle $3\alpha$ configuration is constructed by 
rotating the $\alpha_1$ and $\alpha_2$ clusters with the angle $\pi/2-\theta$ 
around $-\bvec{D}/3$.
Consequently, an E-$3\alpha$ wave function used here is specified by 
parameters, 
$r$, $D$, $\theta$, and $\Lambda$.
We denote the E-$3\alpha$ wave function
$\Phi_{\textrm{E-}3\alpha}(r,D,\theta,\Lambda)$, which is expressed by 
a single Slater determinant.

\begin{figure}[!h]
\begin{center}
\includegraphics[width=7.5cm]{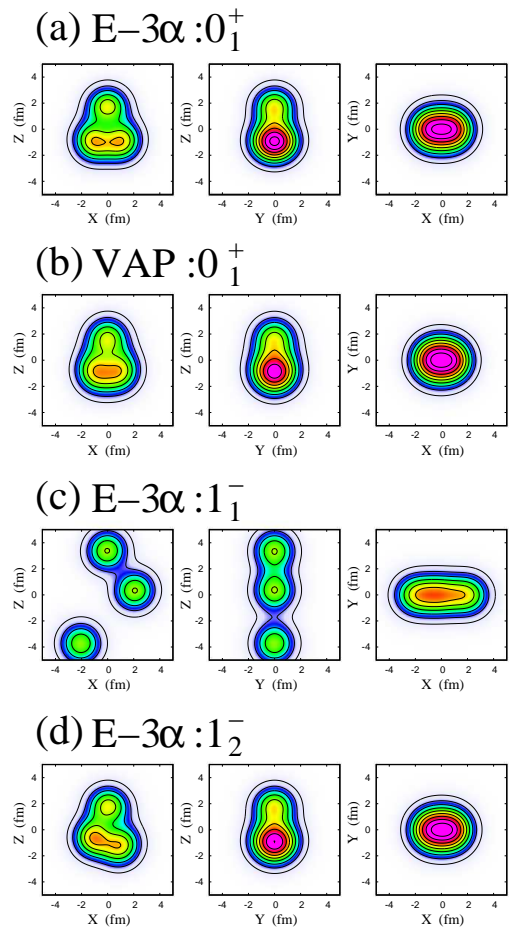} 	
\end{center}
  \caption{(color online) Intrinsic density distributions of the dominant configurations
written by the E-$3\alpha$ wave functions for
(a) the $0^+_1$, (c) $1^-_1$, and (d) $1^-_2$ states. The intrinsic density of the AMD wave function 
$\Phi_{\rm AMD}({\bvec{Z}^0_\textrm{VAP}})$ obtained by the VAP for 
the ground state is also shown.
\label{fig:c12-dense}}
\end{figure}

We search for the E-$3\alpha$  wave functions that have large overlap 
with the obtained $0^+_1$, $1^-_1$, and $1^-_2$ states
and regard them as dominant configurations of the corresponding 
states. The intrinsic density distributions of the dominant configurations for the 
 $0^+_1$, $1^-_1$, and $1^-_2$ states are shown in Fig.~\ref{fig:c12-dense}.
The  $0^+_1$ state  has 85\% overlap with the $0^+$-projected 
$\Phi_{\textrm{E-}3\alpha}$ with $r=1$ fm, $D=2$ fm, $\theta=\pi/2$, and $\Lambda=0.5$
corresponding to a compact isosceles triangle structure 
with the significant $\alpha$-breaking component. The E-$3\alpha$ wave function has 
almost the consistent structure with the AMD wave function obtained by the VAP for 
the $0^+_1$.
The $1^-_1$ state has 60\% overlap with the BB-cluster $3\alpha$ wave function with $r=5.8$ fm, $D=5$ fm,
and $\theta=\pi/4$ projected onto the $J^\pi K=1^- 1$ state, indicating the spatially developed $3\alpha$ structure.
The $1^-_2$ state is the toroidal mode originally obtained as the $1^-_1$ state 
in the sAMD calculation. It has large overlap with
$\Phi_{\textrm{E-}3\alpha}$ with $r=1$ fm, $D=2$ fm, $\theta=3\pi/8$, and $\Lambda=0.5$ 
projected onto $J^\pi K=1^- 1$ state, and corresponds to 
the compact triangle configuration with the cluster breaking. It has  75\% overlap  
with the 1p-1h dominant TD mode of the sAMD. For the $1^-_2$ state of the sAMD+GCM, 
the overlap is reduced to be 50\% because of the coupling with 
the large-amplitude $3\alpha$-cluster mode. 
As shown in Fig.~\ref{fig:c12-dense}, the dominant configuration 
of the TD mode for the $1^-_2$ state shows the intrinsic structure 
similar to that for the $0^+_1$ state except for the slight rotation (tilting) of the $2\alpha$ part. 
\begin{figure}[!h]
\begin{center}
\includegraphics[width=7.5cm]{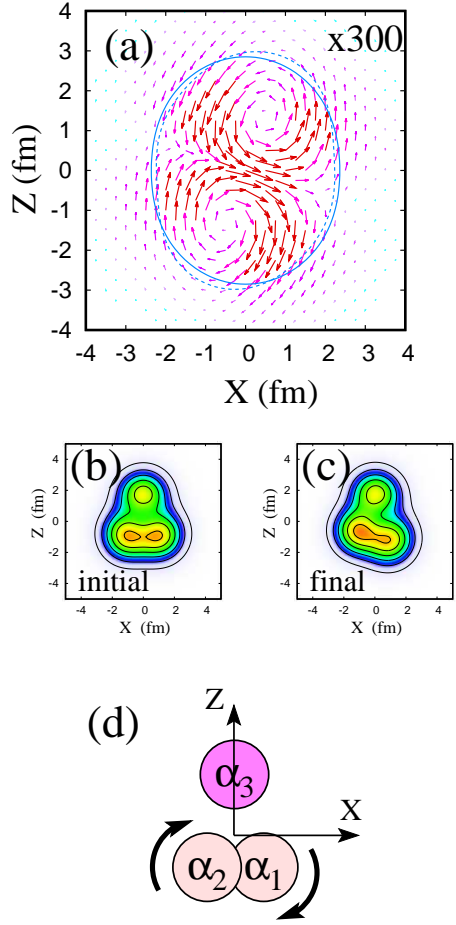} 	
\end{center}
  \caption{(color online) (a)
Transition current density 
for $0^+_1 \to 1^-_2$ in the intrinsic frame (at $Y=0$ on the $X$-$Z$ plane).
The vector plot of 
the transition current density for the parity-projected states of 
the dominant E-$3\alpha$ configurations
is shown. Red solid and magenta dashed lines indicate 
contours for the matter densities 
$\rho(X,0,Z)=0.08$ fm$^{-3}$ of the parity-projected initial and final states, respectively.
(b) Intrinsic density distributions of the dominant configuration 
for the initial state ($0^+_1$) and (c) that of the final state ($1^-_2$). 
(d) Schematic figure for the $0^+_1 \to 1^-_2$  excitation in the cluster picture.
$\alpha_1$ and $\alpha_2$ clusters contain the cluster breaking. 
\label{fig:c12-current}}
\end{figure}

Figure \ref{fig:c12-current} shows
the transition current density for the transition between the dominant 
configurations of the $0^+_1$ and $1^-_2$. 
The transition current density from the positive-parity projected initial state to the 
negative-parity projected final state is shown in Fig.~\ref{fig:c12-current}(a). 
The intrinsic densities without
the parity projection of the initial and final configurations are shown again 
in Figs.~\ref{fig:c12-current}(b) 
and (c).  The transition current density clearly shows
the toroidal nuclear current with the $K=1$ feature in the prolately deformed intrinsic system, 
and describes the remarkable TD strengths for the $0^+_1\to
1^-_2$ transition. In a cluster picture, the toroidal current is understood by the rotation of the 
$2\alpha$ subsystem as shown in a schematic figure of Fig.~\ref{fig:c12-current}(d). 
It means that the $2\alpha$-cluster rotation 
induces the TD dominant dipole excitation. 
Note that the  $2\alpha$ subsystem in the initial and final states consists of  
not ideal $\alpha$ clusters but somewhat dissociated ones 
expressed by the $\alpha$-breaking parameter $\Lambda=0.5$ as mentioned previously. 

Similar mechanism was found in $^{10}$Be as discussed in Ref.~\cite{Kanada-Enyo:2017uzz}. 
In the case of $^{10}$Be, the system is approximately understood by the
 $^6\textrm{He}+\alpha$ clustering, and the TD dominant dipole excitation 
in the $1^-_1$ state is described by rotation of the $^6\textrm{He}$ cluster.
An interesting difference from $^{10}$Be is that 
the large-amplitude cluster mode exists below the TD mode 
in the dipole excitations of $^{12}$C. 

\subsection{TD mode in shell-model limit}
As discussed previously, the TD mode 
can be understood as the rotational excitation of 
the $2\alpha$ subsystem. Instead of the cluster picture, it is worth to discuss 
features of the TD mode in terms of 1p-1h excitations. Indeed, 
the $1^-_2$ contains 40\% $1\hbar\omega$ component 
as the dominant configuration, though it has 
significant mixing of higher shell components because of the coupling with the $3\alpha$-cluster state.
We here consider 1p-1h representation of the TD mode by taking 
a shell-model limit of the E-$3\alpha$ wave functions for the dominant configurations. 

As explained previously, 
the  ground state is approximately described by $\Phi_{\textrm{E-}3\alpha}$
with $r=1$ fm, $D=2$ fm, $\theta=\pi/2$, $\Lambda=0.5$ projected onto
the $J^\pi=0^+$ state,
whereas the dominant component of the TD mode is described by
$\Phi_{\textrm{E-}3\alpha}$
with $r=1$ fm, $D=2$ fm, $\theta=3\pi/8$, $\Lambda=0.5$ projected onto 
the $J^\pi K=1^- 1$ state.
By taking the limit of $\sqrt{\nu}r << 1$ and $\sqrt{\nu}D << 1$, 
we can map the dominant configurations to $0\hbar$ and $1\hbar\omega$ configurations 
in the shell-model limit for the ground state and the TD mode, 
respectively. 
For simplicity, we discuss the shell-model limit
of the $K^\pi$-projected configuration in the intrinsic frame. 

The shell-model limit of the E-$3\alpha$ wave function
is represented by four-nucleon (4N:two protons and two neutrons) configurations
around the SU(3)-limit 2$\alpha$ core with the
$|000\rangle^4 |001\rangle^4$ configuration. Here 
$|n_xn_yn_z\rangle$ means the single-particle state 
in the harmonic oscillator potential with the
size parameter $b=1/\sqrt{2\nu}$.
For single-particle states around the $|000\rangle^4 |001\rangle^4$ core, 
we use the following 
notations,
\begin{eqnarray}
|p_{+\frac{3}{2}}\rangle&=&\frac{1}{\sqrt{2}}\left( |100\rangle+i|010\rangle \right ) |\uparrow \rangle,\\
|p_{+\frac{1}{2}}\rangle&=&\frac{1}{\sqrt{2}}\left( |100\rangle+i|010\rangle \right )|\downarrow \rangle,\\
|p_{-\frac{1}{2}}\rangle&=& \frac{1}{\sqrt{2}}\left( |100\rangle-i|010\rangle \right )|\uparrow \rangle,\\
|p_{-\frac{3}{2}}\rangle&=& \frac{1}{\sqrt{2}}\left( |100\rangle-i|010\rangle \right ) | \downarrow\rangle, \\
|sd_{+\frac{1}{2}}\rangle&=& |002\rangle |\uparrow \rangle,\\
|sd_{-\frac{1}{2}}\rangle&=& |002\rangle|\downarrow \rangle.
\end{eqnarray}

For the $K^\pi=0^+$- and $K^\pi=0^-$-projected states 
of the ground state and the TD mode,  
the 4N configurations in the shell-model limit are 
written as 
\begin{widetext}
\begin{eqnarray}
P^+_{K=0} \Phi_{4N}(\textrm{g.s.})&=&\lambda_+^4   |p_{+\frac{3}{2}} 
p_{-\frac{3}{2}} p_{+\frac{3}{2}} p_{-\frac{3}{2}}\rangle |ppnn \rangle\nonumber \\
&+&\lambda_+^2  \lambda_-^2 |p_{+\frac{3}{2}} 
p_{-\frac{3}{2}} p_{+\frac{1}{2}} p_{-\frac{1}{2}}\rangle \left(|ppnn\rangle+|nnpp\rangle+|pnnp\rangle+|nppn\rangle\right)\nonumber \\
&+&\lambda_-^4   |p_{+\frac{1}{2}} 
p_{-\frac{1}{2}} p_{+\frac{1}{2}} p_{-\frac{1}{2}}\rangle |ppnn\rangle, 
\label{eq:limit-gs}\\
P^-_{K=1} \Phi_{4N}(\textrm{TD})&=&
\lambda_+^4 \left (  |p_{+\frac{3}{2}} sd_{-\frac{1}{2}} p_{+\frac{3}{2}} p_{-\frac{3}{2}}\rangle 
+ |p_{+\frac{3}{2}} p_{-\frac{3}{2}} p_{+\frac{3}{2}} sd_{-\frac{1}{2}}\rangle \right)
 |ppnn  \rangle\nonumber \\
&+&\lambda_+^2  \lambda_-^2 \left (|p_{+\frac{3}{2}} 
sd_{-\frac{1}{2}} p_{+\frac{1}{2}} p_{-\frac{1}{2}}\rangle +  |p_{+\frac{3}{2}} 
p_{-\frac{3}{2}} p_{+\frac{1}{2}} sd_{+\frac{1}{2}}\rangle \right) \left(|ppnn\rangle+|nnpp\rangle+|pnnp\rangle+|nppn\rangle\right)\nonumber \\
&+&\lambda_-^4  \left (|p_{+\frac{1}{2}} sd_{+\frac{1}{2}} p_{+\frac{1}{2}} p_{-\frac{1}{2}}\rangle 
+ |p_{+\frac{1}{2}} p_{-\frac{1}{2}} p_{+\frac{1}{2}} sd_{+\frac{1}{2}}\rangle \right) |ppnn\rangle, \label{eq:limit-td}
\end{eqnarray}
\end{widetext}
where $\lambda_+:\lambda_-=(1+\Lambda):(1-\Lambda)$.
For simplicity, we here assume 
$r/\sqrt{\nu} << D/\sqrt{\nu} <<1$ and an enough small $\theta$
to omit spin-flip excitations.
It is shown that the shell-model
limit for the ground state is not a simple paring state but 
linear combination of 
correlating $nn$, $pp$, and $np$ pairs in $p_{+3/2}p_{-3/2}$
and $p_{+1/2}p_{-1/2}$. 
Spin and isospin configurations are strongly correlated with each other 
because of the $\alpha$-like correlation.
The coefficients of the $p_{+3/2}p_{-3/2}p_{+3/2}p_{-3/2}$,  $p_{+3/2}p_{-3/2}p_{+3/2}p_{-3/2}$, and $p_{+1/2}p_{-1/2}p_{+1/2}p_{-1/2}$
terms are determined by the $\alpha$-breaking parameter $\Lambda$.
The  $\Lambda=1$ case corresponds to the uncorrelated $4N$ state 
with the pure $p_{+3/2}p_{-3/2}p_{+3/2}p_{-3/2}$ configuration around the $2\alpha$ core.
The TD mode is expressed by 1p-1h excitations of 
$p^{-1}_{-3/2}sd_{-1/2}$ and $p^{-1}_{-1/2}sd_{+1/2}$
on the top of the ground state configuration.
They are coherent 1p-1h excitations changing 
the oscillator quanta $n_\perp\equiv n_x+n_y$ and $n_z$
as $\Delta n_\perp=-1$ and $\Delta n_z=+2$, and 
contribute to the remarkably strong TD transition. 

\section{Summary and outlook}\label{sec:summary}
We have investigated cluster and toroidal natures of the ISD 
excitations in $^{12}$C based on the sAMD+GCM calculation. 
In the $E=10-15$ MeV region, we have found two LED modes. 
One is the spatially developed $3\alpha$-cluster state and the other is the TD mode.
The TD mode is dominantly described by coherent 1p-1h excitations on the 
ground state. The cluster state comes down to the energy lower than the TD mode 
because of the large amplitude cluster motion.

For the CD (ordinary ISD) excitations, 
the transition strengths
are mainly distributed in the high-energy region for the ISGDR, whereas 5\% of the TEWS 
exist in $E<20$ MeV consistently to the experimental data. 
In the experimental data of the CD strengths, 
the bump structure at $E\sim 15$ MeV is a candidate for the $1^-_2$ state 
of the 1p-1h dominant TD mode. 
The LED states are strongly excited by the TD
operator. In the present calculation, two modes, the cluster and TD modes, are coupled with each other. 
As a result of the coupling, the TD strengths for the $1^-_1$ and $1^-_2$ states are the same order.  
Unfortunately, there is no established method to experimentally 
observe the TD strengths.

We have discussed the feature of the TD mode from the cluster picture 
based on the model analysis by analyzing dominant configurations
in the single-Slater expression. 
In the analysis, the transition current density clearly shows 
the toroidal nature of the TD mode induced by the rotation of the $2\alpha$-like 
deformed subsystem. We have also discussed the connection 
of the rotational excitation of the cluster with 1p-1h excitations
by taking the shell-model limit of the dominant configurations.
The TD mode can be understood as the coherent 1p-1h excitations 
on the ground state, in which spin and isospin configurations are 
highly correlated because of the $\alpha$-type four-body 
correlation.

\begin{acknowledgments}
The computational calculations of this work were performed by using the
supercomputer in the Yukawa Institute for theoretical physics, Kyoto University. This work was supported by 
JSPS KAKENHI Grant Nos. 26400270 and 16J05659.
\end{acknowledgments}

\end{document}